\documentclass[journal]{IEEEtran}
\usepackage{amsfonts}
\usepackage{cite}
\usepackage{graphicx}
\usepackage{amsmath}
\usepackage{amssymb}
\usepackage{algorithm}
\usepackage{algorithmic}

\begin{document}

\title{Joint LDPC and Physical-layer Network Coding for Asynchronous Bi-directional Relaying}

\author{Xiaofu~Wu, ~Chunming~Zhao,
        and Xiaohu~You 
\thanks{This work was supported in part by the National Science
        Foundation of China under Grants 60972060, 61032004. The work of X. Wu was also supported by the National Key S\&T Project under Grant 2010ZX03003-003-01 and by the Open Research Fund of National Mobile Communications Research Laboratory, Southeast University (No. 2010D03).
        }
\thanks{Xiaofu Wu is with the Nanjing Institute of Communications
        Engineering, Nanjing 210007, China. He is also with the National Mobile Commun. Research Lab., Southeast Univ., Nanjing
        210096 (Email: xfuwu@ieee.org).}
\thanks{Chunming Zhao and Xiaohu You are with the National Mobile Commun. Research Lab.,
        Southeast University, Nanjing 210096, China (Email: cmzhao@seu.edu.cn, xhyu@seu.edu.cn).}}


\maketitle

\begin{abstract}
In practical asynchronous bi-directional relaying, symbols transmitted by two sources cannot arrive at the relay with perfect frame and symbol alignments and the asynchronous multiple-access channel (MAC) should be seriously considered. Recently, Lu et al. proposed a Tanner-graph representation of the symbol-asynchronous MAC with rectangular-pulse shaping and further developed the message-passing algorithm for optimal decoding of the symbol-asynchronous physical-layer network coding.  In this paper, we present a general channel model for the asynchronous MAC with arbitrary pulse-shaping. Then, the Bahl, Cocke, Jelinek, and Raviv (BCJR) algorithm is developed for optimal decoding of the asynchronous MAC channel. For Low-Density Parity-Check (LDPC)-coded BPSK signalling over the symbol-asynchronous MAC, we present a formal log-domain generalized sum-product-algorithm (Log-G-SPA) for efficient decoding. Furthermore, we propose to use cyclic codes for combating the frame-asynchronism and the resolution of the relative delay inherent in this approach can be achieved by employing the simple cyclic-redundancy-check (CRC) coding technique. Simulation results demonstrate the effectiveness of the proposed approach.
\end{abstract}


\IEEEpeerreviewmaketitle

\section{Introduction}
\PARstart{N}{etwork} coding has shown its power for disseminating information over networks \cite{KoetterAlg,Chou}.
For wireless cooperative networks, there are increased interests in employing the idea of network
coding for improving the throughput of the network.
Indeed, the gain is very impressive for the special bi-directional relaying
scenarios with two-way or multi-way traffic as addressed in \cite{Popovski}.

\begin{figure}[htb]
   \centering
   \includegraphics[width=0.45\textwidth]{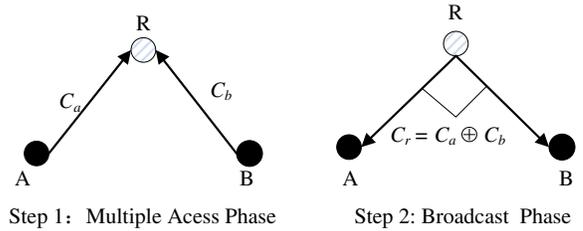}
   \caption{Bi-directional relaying with PNC.}
   \label{fig:sys}
\end{figure}

For bi-directional relaying, two sources A and B want to exchange information with each other by the help of a relay node R as shown in Fig. \ref{fig:sys}. Traditionally, this can be achieved via four steps. Recently, it was recognized that only two steps are essentially required with the employment of the powerful idea of physical-layer network coding (PNC)\cite{Zhang_PLNC}. In particular, the superimposed signal received at the relay can be viewed as the physically-combined network coding form of the two source messages further impaired by the channel noise. Hence, the so-called physical-layer network coding  can be well employed to improve the throughput of bi-directional relaying.

For bi-directional relaying with PNC,  it is assumed that communication takes place in two phases - a multiple
access phase and a broadcast phase as shown in Fig. \ref{fig:sys}. In the first phase, the two source nodes send signals simultaneously to the relay.
In the second phase, the relay processes the superimposed signal of the simultaneous packets and
maps them to a network-coded (XOR) packet for broadcast back to the source nodes. Then, both sources can retrieve their own information  as they know completely what they have sent. Compared with the traditional relay system, PNC doubles the throughput of
the two-way relay channel.

To be more practical, channel coding should be employed to further improve the reliability of the system. In \cite{ZhangPNC,ZhangJSAC}, joint channel decoding and physical layer network coding (JCNC) have been introduced. It was recognized that with the same linear channel code at both source nodes, the XOR of both source codewords is still  a valid codeword. Thus, the received signal can be decoded to the XOR of the source information at the relay without changing the decoding algorithm. In \cite{WuWeb}, we derived the closed-form expression for computing the log-likelihood ratios (LLRs) of the network-coded codeword for a complex multiple-access channel, and it was revealed that the equivalent channel observed at the relay is an asymmetrical channel. Although this approach can be efficiently implemented, it does result in performance loss due to the use of $\Pr(c_a(k)\oplus c_b(k)|\mathbf{r}_k)$ \footnotemark\footnotetext{For the proper definition, we refer readers to Section-II.} while the joint probabilities $\Pr(c_a(k),c_b(k)|\mathbf{r}_k)$ is not fully used. In \cite{ZhangJSAC}, a novel decoding scheme based on the arithmetic-sum of the source codewords was proposed for repeat-accumulate (RA) codes. A generalized sum-product algorithm (G-SPA) over the Galois field $GF(2^2)$ was proposed in \cite{WubbenBPSK} for LDPC coded BPSK system, which can work directly with the joint probabilities  $\Pr(c_a(k),c_b(k)|\mathbf{r}_k)$ and a significant gain was observed compared to the JCNC approach. Its extension to QPSK signalling was developed in \cite{WubbenQPSK}.

A key issue in practical PNC is how to deal with the asynchrony between the signals transmitted
by the two source nodes. That is, symbols transmitted by the two source nodes
could arrive at the receiver with both symbol and frame misalignments.

In \cite{LuWeb}, Lu et al. proposed a Tanner-graph representation of the symbol-asynchronous multiple-access channel (MAC) with rectangular-pulse shaping  and also developed the message-passing algorithm for optimal decoding of asynchronous physical-layer network coding. Furthermore, the message-passing algorithm was also developed when RA codes are employed. This message-passing algorithm is, in essence, the cascade of the BCJR algorithm for the asynchronous physical-layer network coding and the G-SPA \cite{WubbenBPSK} over the underlying Tanner-graph of the specified RA code. However, its current form is in the probability-domain with channel coding restricted to the special RA codes, which is not desirable in practice.

In this paper, we provide further insights into the asynchronous bi-directional relaying. In particular, the general asynchronous MAC channel with arbitrary pulse-shaping is developed and its connection to the rectangular-pulse shaping \cite{LuWeb} is discussed. Then, the BCJR formulation of the asynchronous MAC channel is proposed, which can shed lights for various practical algorithms suitable for implementation.  For LDPC-coded BPSK signalling over the symbol-asynchronous MAC, we present a formal log-domain generalized sum-product-algorithm (Log-G-SPA) for efficient decoding. Furthermore, we propose to use cyclic codes for combating the frame-asynchronism and its related problem of delay resolution is discussed in detail.

The rest of the paper is organized as follows. In Section-II, a general channel model for asynchronous physical-layer network coding is developed.
We then formulate the Log-G-SPA decoding for joint LDPC and PNC over asynchronous MACs in Section-III. Section-IV address the problem of frame asynchronism. Simulation results are provided in Section-V, and the conclusion is made in Section-VI.

\section{General Channel Model for Asynchronous Physical-layer Network Coding}
\subsection{Asynchronous Multiple Access Channel Model}

During the multiple-access phase, the source nodes A and B transmit the modulated signals $x_a(t)$ and $x_b(t)$ to the relay simultaneously.
For a general continues-time multiple-access channel, the received signal at the relay can be
expressed as
\begin{eqnarray}
 \label{eq:c1}
    y(t) &=& h_a x_a(t)+ h_b x_b(t)+w(t) \nonumber \\
           &=& \sum_{k=0}^{\infty} h_a c_a(k)g_a(t-k T - \tau_a) \nonumber \\
              & &+ \sum_{k=0}^{\infty} h_b c_b(k) g_b(t-k T - \tau_b) + w(t),
\end{eqnarray}
where the delays $\tau_a \in [0,T), \tau_b \in [0,T)$ account for the symbol asynchronism between source nodes A and B, $w(t)$ is the complex white Gaussian noise with power spectral density equal to $\frac{\sigma^2}{2}$, the channel coefficients $h_a, h_b$ are complex channel gains keeping fixed during transmission, $\{c_a(k)\},\{c_b(k)\}$ are the modulated sequences, and $g_a(t), g_b(t)$ are normalized pulse-shaping functions ($\int_{-\infty}^{\infty} |g_a(t)|^2 dt = 1$) for source nodes A and B, respectively. In this section, we focus on the symbol asynchronism. Without loss of generality, we assume that $0\leq \tau_a \leq \tau_b<T$ and both of them are known to the receiver. The frame asynchronism is considered in section-IV.

By passing the observations through two matched filters for signals $x_a(t)$ and $x_b(t)$, respectively, one can get the following discrete-time samples
\begin{eqnarray}
 \label{eq:d1}
    y_a(k) &=&  \int_{-\infty}^{\infty} y(t)g_a^*(t-kT-\tau_a)dt, \nonumber \\
    y_b(k) &=&  \int_{-\infty}^{\infty} y(t)g_b^*(t-kT-\tau_b)dt.
\end{eqnarray}

It can be well understood that the discrete samples $\left\{[y_a(k), y_b(k)]^T\right\}$ are sufficient statistics for the maximum $a$ $posteriori$ (MAP) symbol detection as explained in \cite{VerduIT89}. By incorporating (\ref{eq:c1}) into (\ref{eq:d1}), it follows that
\begin{eqnarray}
 \label{eq:d2}
    y_a(k) &=& h_a c_a(k) + \sum_l h_b \rho_{ab}(l) c_b(k-l) + w_a(k), \nonumber \\
    y_b(k) &=& h_b c_b(k) + \sum_l h_a \rho_{ba}(l) c_a(k-l) + w_b(k),
\end{eqnarray}
where
\begin{eqnarray}
 \label{eq:d3}
    \rho_{ab}(l) &=&  \int_{-\infty}^{\infty} g_a^*(t) g_b(t+lT+\tau_a-\tau_b) dt, \nonumber \\
    \rho_{ba}(l) &=& \int_{-\infty}^{\infty} g_b^*(t) g_a(t+lT+\tau_b-\tau_a) dt,
\end{eqnarray}
and
\begin{eqnarray}
 \label{eq:d4}
    w_a(k) &=&  \int_{-\infty}^{\infty} w(t)g_a^*(t-kT-\tau_a)dt, \nonumber \\
    w_b(k) &=&  \int_{-\infty}^{\infty} w(t)g_b^*(t-kT-\tau_b)dt.
\end{eqnarray}

One can also rewrite (\ref{eq:d2}) in the matrix form, as shown at the top of the next page.
\begin{figure*}[!t]
\begin{eqnarray}
\label{eq:A3}
\left[\begin{array}{c}
y_a(k) \\
y_b(k)
\end{array}\right] = \left[\begin{array}{cc}
h_a & h_b\rho_{ab}(0) \\
h_a\rho_{ba}(0) & h_b
\end{array}\right]
\left[\begin{array}{c}
c_a(k) \\
c_b(k)
\end{array}\right] 
 +  \sum_l \left[\begin{array}{cc}
0 & h_b \rho_{ab}(l) \\
h_a \rho_{ba}(l) & 0
\end{array}\right]
\left[\begin{array}{c}
c_a(k-l) \\
c_b(k-l)
\end{array}\right] +
\left[\begin{array}{c}
w_a(k) \\
w_b(k)
\end{array}\right].
\end{eqnarray}
\centering
\end{figure*}

Here, the discrete random process $\left\{[w_a(k),w_b(k)]^T\right\}$ is Gaussian with zero mean and covariance matrix:
\begin{eqnarray}
\label{eq:noise1}
\frac{1}{2} E\left[\left[\begin{array}{c}
w_a(k) \\
w_b(k)
\end{array}\right] \cdot
\left[\begin{array}{cc}
w_a^*(j), w_b^*(j)
\end{array}\right]\right] = \sigma^2 \mathbf{\Lambda}(k-j)
\end{eqnarray}
where $\mathbf{\Lambda}(k)=0$ if $|k|>L$ and $\mathbf{\Lambda}(0)$, $\mathbf{\Lambda}(l)$, $\mathbf{\Lambda}(-l)$ are given as follows
\begin{eqnarray}
    \label{eq:cor0}
    \mathbf{\Lambda}(0) = \left[\begin{array}{cc}
    1 & \rho_{ab}(0) \\
    \rho_{ba}(0) & 1
\end{array}\right],
\end{eqnarray}
\begin{eqnarray}
    \label{eqn:cor1}
    \mathbf{\Lambda}(l) = \mathbf{\Lambda}^\dag(-l) = \left[\begin{array}{cc}
    0 & \rho_{ba}(l) \\
    \rho_{ab}(l) & 0
\end{array}\right],  l=1,\cdots,L.
\end{eqnarray}
Here, $L$ denotes the memory length of the channel, which is determined by the correlation of the pulse-shaping functions (\ref{eq:d3}).

For convenience of the MAP detection, the whitened matched filter (WMF) is often employed for transforming the received signal into a discrete
time sequence with minimum-phase channel response and white noise. This procedure often simplifies analysis and is a first step in the implementation of some estimators, including the maximum-likelihood sequence estimation
detector (MLSE) and the MAP detector. The WMF is determined by factoring the channel
spectrum into a product of a minimum phase filter and its time inverse.

Let $\mathbf{\Omega}(z)$ be the (two-sided) $z$ transform of the sampled autocorrelation sequences $\mathbf{\Lambda}(k)$, i.e.,
\begin{eqnarray}
    \label{eqn:zf4}
    \mathbf{\Omega}(z) = \sum_{k=-L}^L \mathbf{\Lambda}(k)z^{-k}.
\end{eqnarray}
By noting the property (\ref{eqn:cor1}), it follows that ${\Omega}(z)$ can be factored as
\begin{eqnarray}
    \label{eqn:factor}
    \mathbf{\Omega}(z) = \mathbf{F}^\dag(z^{-1}) \mathbf{F}(z).
\end{eqnarray}
By invoking the spectral factorization theorem, it is reasonable to find a physically realizable, stable discrete-time filter $\left(\mathbf{F}^\dag(z^{-1})\right)^{-1}$, which can transform a colored random process into a white random process.

In \cite{Duel}, it has been shown that  $\mathbf{F}(z)$ has the form of
\begin{eqnarray}
    \label{eqn:f1}
    \mathbf{F}(z) = \sum_{l=0}^L F_l z^{-l},
\end{eqnarray}
where
\begin{eqnarray}
    \label{eqn:f1}
     F_l = \left[\begin{array}{cc}
    f_{aa}^l & f_{ab}^l \\
    f_{ba}^l & f_{bb}^l
\end{array}\right].
\end{eqnarray}

Consequently, passage of the received vector sequence $\left\{\mathbf{y}(k)=[y_a(k),y_b(k)]^T\right\}$ through the digital filter $\left(\mathbf{F}^\dag(z^{-1})\right)^{-1}$ results into an output vector sequence $\left\{\mathbf{r}(k)\right\}$ that can be expressed at the top of the next page. Now, the discrete random process $\left\{\mathbf{n}(k)=[n_a(k),n_b(k)]^T\right\}$ is  zero-mean white Gaussian process with covariance of $\sigma^2 \mathbf{I}$.
\begin{figure*}[!t]
\begin{eqnarray}
\label{eq:W}
\left[\begin{array}{c}
r_a(k) \\
r_b(k)
\end{array}\right] =  \sum_{l=0}^L \left[\begin{array}{cc}
h_a f_{aa}^l & h_b f_{ab}^l \\
h_a f_{ba}^l & h_b f_{bb}^l
\end{array}\right]
\left[\begin{array}{c}
c_a(k-l) \\
c_b(k-l)
\end{array}\right] 
 +
\left[\begin{array}{c}
n_a(k) \\
n_b(k)
\end{array}\right].
\end{eqnarray}
\centering
\end{figure*}

Let $\mathbf{c}_{ab}(k)=\left[\begin{array}{c}
c_a(k) \\
c_b(k)
\end{array}\right]$, and $\mathbf{r}(k)=\left[\begin{array}{c}
r_a(k) \\
r_b(k)
\end{array}\right]$. Then, the formulation (\ref{eq:W}) can be elegantly expressed as
\begin{eqnarray}
 \label{eq:d4}
    \mathbf{r}(k) &=& \Psi \left(\mathbf{c}_{ab}(k), \cdots, \mathbf{c}_{ab}(k-L)\right) + \mathbf{n}(k) \nonumber \\
                  &\triangleq& \Psi \left(\{\mathbf{c}_{ab}(k-l)\}_{l=0}^L\right) + \mathbf{n}(k).
\end{eqnarray}
It is clear that the function $\Psi(\cdot,\cdot)$ is linear. By assuming the ideal knowledge on $\Psi(\cdot,\cdot)$ and $\sigma^2$, the asynchronous MAC can be modeled as the vector inter-symbol interference (ISI) channel. To estimate the a $posteriori$ probability (APP) $\Pr\left(\mathbf{c}_{ab}(k)|\mathbf{r}_0^{N-1}\right)$, the BCJR algorithm can be naturally employed. Here, $N$ denotes the observation length at the relay.

\subsection{Rectangular-pulse shaping}
Let $\delta = \frac{\tau_b-\tau_a}{T}$ denote the relative delay between source nodes A and B. For the rectangular pulse-shaping functions $g_a(t), g_b(t)$, i.e., $g_a(t)=g_b(t)=u(t)-u(t-T)$ with $u(t)$ denoting the unit step function, the authors in \cite{LuWeb} proposed to consider the following discrete-time samples
\setlength{\arraycolsep}{0.0em}
\begin{eqnarray}
 \label{eq:r1}
    y_e(k) &=& \frac{1}{\delta } \int_{kT+\tau_a}^{kT+\tau_b} y(t)g(t-kT-\tau_a)dt  \nonumber \\
    y_o(k) &=& \frac{1}{(1-\delta)} \int_{kT+\tau_b}^{(k+1)T+\tau_a} y(t)g(t-kT-\tau_a)dt.
\end{eqnarray}
It is clear that the matched-filter outputs (\ref{eq:d1}) can be well related to (\ref{eq:r1}) as follows:
\begin{eqnarray}
     y_a(k) &=& \delta y_e(k)+ (1-\delta) y_o(k), \nonumber  \\
     y_b(k) &=& (1-\delta)y_o(k)+\delta y_e(k+1).
\end{eqnarray}
Hence, the samples $\left\{[y_e(k), y_o(k)]^T\right\}$ are also the sufficient statistics for the MAP detection.
By combining (\ref{eq:c1}) and (\ref{eq:r1}), it follows that
\begin{eqnarray}
 \label{eq:r2}
    y_e(k) &=& h_a c_a(k) + h_b c_b(k-1) + w_e(k) \nonumber \\
    y_o(k) &=& h_a c_a(k) + h_b c_b(k) + w_o(k),
\end{eqnarray}
\setlength{\arraycolsep}{5pt}
where $w_e(k)$ and $w_o(k)$ are independent zero-mean complex Gaussian variables with variance of $\frac{1}{\delta} \sigma^2$ and $\frac{1}{1-\delta} \sigma^2$. Hence, one can write (\ref{eq:r2}) as the following matrix form

\begin{eqnarray}
\label{eq:r3}
\left[\begin{array}{c}
y_e(k) \\
y_o(k)
\end{array}\right] &=& \left[\begin{array}{cc}
0 & h_b \\
0 & 0
\end{array}\right]
\left[\begin{array}{c}
c_a(k-1) \\
c_b(k-1)
\end{array}\right] \nonumber \\
&+&  \left[\begin{array}{cc}
h_a & 0 \\
h_a & h_b
\end{array}\right]
\left[\begin{array}{c}
c_a(k) \\
c_b(k)
\end{array}\right]
+
\left[\begin{array}{c}
w_e(k) \\
w_o(k)
\end{array}\right].
\end{eqnarray}
Hence, the equivalent ISI channel model (\ref{eq:W}) is still valid.

\subsection{BCJR Algorithm}
In this subsection, we formulate the BCJR algorithm \cite{BCJR}, which is known to be optimal in implementing the
MAP symbol detection for linear channels with finite memory.

Let us define, at time epoch $k$, the state $s_k$ as
\begin{equation}
\label{eq:sd}
   s_k=\left(\mathbf{c}_{ab}(k-1),\cdots,\mathbf{c}_{ab}(k-L)\right)
\end{equation}
and the branch metric function as
\setlength{\arraycolsep}{0pt}
\begin{eqnarray}
\label{eq:sd}
   \gamma_k && (s_k,\mathbf{c}_{ab}(k)) \propto \Pr(\mathbf{c}_{ab}(k)) \nonumber   \\
    &&  \cdot \exp\left(-\frac{\left|\mathbf{r}(k)- \Psi \left(\{\mathbf{c}_{ab}(k-l)\}_{l=0}^L\right)\right|^2}{2\sigma^2}\right).
\end{eqnarray}
\setlength{\arraycolsep}{5pt}
The BCJR algorithm is characterized by the following forward
and backward recursions:
\begin{eqnarray}
\label{eq:fd}
   \alpha_{k+1}(s_{k+1})=\sum_{\mathbf{c}_{ab}(k)} \sum_{s_k} \mathcal{T}(\mathbf{c}_{ab}(k),s_k,s_{k+1}) \nonumber \\
    \cdot \alpha_k(s_k) \gamma_k(s_k,\mathbf{c}_{ab}(k)),
\end{eqnarray}
where $\mathcal{T}(\mathbf{c}_{ab}(k), s_k, s_{k+1})$ is the trellis indicator function, which is equal to 1 if $\mathbf{c}_{ab}(k), s_k$, and $s_{k+1}$
satisfy the trellis constraint and 0 otherwise;

\begin{eqnarray}
\label{eq:fd}
   \beta_{k}(s_k)=\sum_{\mathbf{c}_{ab}(k)} \sum_{s_{k+1}} \mathcal{T}(\mathbf{c}_{ab}(k),s_k,s_{k+1}) \nonumber \\
   \cdot \beta_{k+1}(s_{k+1}) \gamma_k(s_k,\mathbf{c}_{ab}(k)).
\end{eqnarray}
Then, the joint APPs $\Pr\left(\mathbf{c}_{ab}(k)|\mathbf{r}_0^{N-1}\right)$ can be calculated as
\setlength{\arraycolsep}{0.0em}
\begin{eqnarray}
\label{eq:llr}
    \Pr &&\left(\mathbf{c}_{ab}(k)|\mathbf{r}_0^{N-1}\right)  \nonumber \\
    && =\sum_{s_{k+1}}\mathcal{T}(\mathbf{c}_{ab}(k),s_{k+1}) \alpha_{k+1}(s_{k+1}) \beta_{k+1}(s_{k+1}),
\end{eqnarray}
where the indicator function $\mathcal{T}(\mathbf{c}_{ab}(k),s_{k+1})$ is equal to 1 if $s_{k+1}$
is compatible with $\mathbf{c}_{ab}(k)$ and 0 otherwise.

For rectangular-pulse shaping, it should be pointed out that $L=1$ and the value of $\Psi \left(\mathbf{c}_{ab}(k), \mathbf{c}_{ab}(k-1)\right)$ is independent of $c_a(k-1)$, hence the state $s_k$ can be further simplified as $s_k= \left( c_b(k-1) \right)$.

Just like in \cite{Hagenauer}, the proposed BCJR algorithm can be implemented efficiently in the log-domain, i.e., the Log-BCJR algorithm (or the Log-MAP algorithm). In what follows, we denote by $\mathcal{B}_{MAC}(\mathbf{r}_0^{N-1}, L_i(\mathbf{c}_{ab}(k)), L_e(\mathbf{c}_{ab}(k)))$ as the Log-BCJR algorithm, where $L_i(\mathbf{c}_{ab}(k))$ denotes the $a$ $priori$ information and $L_e(\mathbf{c}_{ab}(k)))$ the $a$ $posteri$ $extrinsic$ information, both in the log-domain. The further simplification of the Log-MAP algorithm, such as the Max-Log-MAP algorithm, is also straightforward, with some potential performance loss.

With the joint APPs $\Pr\left(\mathbf{c}_{ab}(k)|\mathbf{r}_0^{N-1}\right)$, one can calculate the APPs of the XOR codeword  $\Pr\left(c_a(k)\oplus c_b(k)|\mathbf{r}_0^{N-1}\right)$ for physical network coding. If both sources A and B assume the same linear channel code, the relay node can make use of $\Pr\left(c_a(k)\oplus c_b(k)|\mathbf{r}_0^{N-1}\right)$ to perform channel decoding to obtain the pairwise XOR of the source symbols. However, this disjoint channel-decoding and network-coding scheme, i.e., the JCNC scheme, performs worse than the joint channel-decoding and network-coding scheme, i.e., the G-SPA scheme \cite{WubbenBPSK, LuWeb}.

\section{Log-G-SPA Decoding of Joint LDPC and PNC over the Asynchronous MAC}
In this paper, the employment of the LDPC coding scheme is assumed for both sources A and B.
Let $C_a$ be a $(N, K_a)$ LDPC code of block length $N$ and dimension
$K_a$ for source A, which has a parity-check matrix $H_a=[h_{m,n}]$ of $M$ rows, and
$N$ columns. Let $R_a=K_a/N$ denote its code rate.
Correspondingly, we can define the code $C_b$ with a parity-check matrix of $H_b$ for source B.

For any given LDPC encoded vector $\mathbf{c}_a = (c_a(0),c_a(1),\cdots, c_a(N-1))^T$ for source A and
$\mathbf{c}_b = (c_b(0),c_b(1),\cdots, c_b(N-1))^T$ for source B, we have
\begin{eqnarray}
 \label{eq:7}
    H_a \mathbf{c}_a &=& \mathbf{0}, \nonumber \\
    H_b \mathbf{c}_b &=& \mathbf{0}.
\end{eqnarray}

For joint LDPC and physical-layer network coding, we consider the employment of the same LDPC code at both sources A and B.
In this case, one have that $H_a=H_b=H\triangleq \left[h_{m,n}\right]$ and
\begin{equation}
 \label{eq:10}
    H (\mathbf{c}_a \oplus \mathbf{c}_b) = \mathbf{0}.
\end{equation}
For the relay R, it tries to decode $\mathbf{c}_r=\mathbf{c}_a \oplus \mathbf{c}_b$.
During the broadcast phase, the relay transmits the XOR codeword $\mathbf{c}_r$ to both sources A and B.
Then, both sources A and B decode $\mathbf{c}_r=\mathbf{c}_a \oplus \mathbf{c}_b$ based on the received signal vector and since they have $\mathbf{c}_a$ and
$\mathbf{c}_b$, they can obtain $\mathbf{c}_b$ and $\mathbf{c}_a$, respectively. Hence, the bottleneck is to decode $\mathbf{c}_r$ for the relay node during the multiple-access phase.

Instead of decoding the source signals separately or by decoding the XOR, the authors in \cite{WubbenBPSK} propose to decode the two codes jointly
with a generalized sum-product algorithm (G-SPA). With this G-SPA decoding, the received superimposed signal is first decoded  to $\{\mathbf{c}_{ab}(k)\}$ with respect to Galois-field $GF(2^2)$ for the BPSK signalling and then the XOR rule is executed before transmission to both sources. This approach almost exploits all available information about the superimposed receive signal as well as the code structure, hence it can achieve excellent performance.

In what follows, we present a log form of the G-SPA (Log-G-SPA) decoding for joint LDPC and physical-layer network coding over the asynchronous multiple-access channel. For convenience, we focus on the BPSK signalling. However, its generalization to the QPSK signalling is straightforward \cite{WubbenQPSK}.

For $H\triangleq \left[h_{m,n}\right]$ and an eligible codeword $(c_0,c_1,\cdots,c_{N-1})$, one have that
\begin{equation*}
 \label{eq:10}
   \sum_n h_{m,n} c_n = 0.
\end{equation*}

For the G-SPA decoding, one can consider a virtual combined encoder which maps the messages generated by both sources A and B into the virtual codeword $\mathbf{c}_{ab}(D)=\left[c^{ab}_0(D),\cdots, c^{ab}_{N-1}(D)\right]$, where $c^{ab}_n(D)=c_a(n) + c_b(n) D$. Let $h_{m,n}^G(D)=1$ if $h_{m,n}=1$, zero otherwise. Hence, each virtual codeword $\mathbf{c}_{ab}(D)$ can be seen as a codeword with elements taken from $GF(2^2)$ and its corresponding parity-check matrix $H^G$ takes values from $GF(2^2)$ with a special constraint of $H^G=H$. Finally, a virtual 4-ary LDPC coding scheme is obtained with the codeword $\mathbf{c}_{ab}(D)$ satisfying
\begin{equation}
 \label{eq:10}
   \sum_n h_{m,n}^G(D) c_n^{ab}(D) = 0  \mod (1+D+D^2).
\end{equation}
This insight can be well employed to develop a generalized SPA over $GF(2^2)$, which is a simpler version of the standard SPA employed in $GF(2^2)$-LDPC coding scheme. Indeed, the permutation step inherent in the standard SPA for decoding of the non-binary LDPC code can be totally neglected thanks to the special form of the parity matrix.

We denote the set of variables that participate in check $m$ by $\mathcal{N}(m)=\{n: h_{m,n}^G(D)=1\}$. Similarly, we denote the set of
checks in which variable $n$ participates as $\mathcal{M}(n)=\{m: h_{m,n}^G(D)=1\}$. We denote by $\mathcal{N}(m)\backslash n$ as the set $\mathcal{N}(m)$ with variable $n$ excluded and by $\mathcal{M}(n)\backslash m$ as the set $\mathcal{M}(n)$ with check $m$ excluded.

We also denote by $V_m$ as the subset of variables corresponding to the non-zero elements in $m$th row of $H^G$, by $GF(4)=\left\{\alpha_0,\alpha_1,\alpha_2,\alpha_3\right\}$ as the finite field of size $4$, by $L(v=\alpha_i)=\ln\left(\Pr(v=\alpha_i|\mathbf{r}_0^N)\right)$ as the log value of the APP $\Pr(v=\alpha_i|\mathbf{r}_0^N)$, by ${L}(v)=[L(v=\alpha_0), L(v=\alpha_1), L(v=\alpha_2), L(v=\alpha_{3})]$ as its vector form. For the Log-G-SPA, the message updated from the variable-to-check message from $n$ to $m$ is denoted by ${L}_{n,m}(v_n)$, while
the check-to-variable message from $m$ to $n$ is denoted by ${L}_{m,n}(v_n)$. The notations of $v_n$ and $\mathbf{c}_{ab}(n)$ can be interchangeably used .

For LDPC coded BPSK signalling over asynchronous MACs, the Log-G-SPA can be formally stated as follows.

\subsection {Log-G-SPA}
\begin{enumerate}
\item[S1:] Initialization:
    \begin{enumerate}
        \item[$I_1$]: A priori information for LDPC decoding   \\ $L_{m,n}(v_n)=0$;

        \item[$I_2$]: A priori information for $\mathcal{B}_{MAC}$ algorithm  \\
        $L_i(\mathbf{c}_{ab}(n))= 0$.
    \end{enumerate}

\item[S2:] Implement the Log-BCJR algorithm for the asynchronous MAC:  \\
        \begin{equation*}
           \mathcal{B}_{MAC}\left(\mathbf{r}_0^N, L_i(\mathbf{c}_{ab}(n)), L_e(\mathbf{c}_{ab}(n))\right),
        \end{equation*}
           and outputs:
        \begin{eqnarray*}
           L_{n,m}(v_n)= L_e(\mathbf{c}_{ab}(n))\triangleq[\ln(\Pr(v_n=\alpha_i|\mathbf{r}_0^N))],
          \end{eqnarray*}
      which is initialized as the variable-to-check messages.
      For first iteration,
      \begin{equation*}
        \tilde{L}_n(v_n)=L_e(\mathbf{c}_{ab}(n)).
      \end{equation*}

\item[S3:] Hard decision:
\begin{eqnarray*}
    \hat{v}_n = \text{arg}\min_{v_n} \tilde{L}_n(v_n):=\hat{c}_a(n)+\hat{c}_b(n)D; \\
    \hat{\mathbf{c}}_r =[\hat{c}_r(0),\cdots,\hat{c}_r(N-1)], \hat{c}_r(n)= \hat{c}_a(n)\oplus \hat{c}_b(n);
\end{eqnarray*}
If $(H  \hat{\mathbf{c}}_r ==\mathbf{0})$ \\
\begin{minipage}{8cm}
\hspace{3ex} output $\hat{\mathbf{c}}_r$ and terminate the decoding;
\end{minipage}

\item[S4:]  Check node processing: \\
\begin{eqnarray*}
\label{eqn:2}
     L_{m,n}(v_n)= \bigoplus_{V_m\backslash v_n}\sum_{n'\in \mathcal{N}(m)\backslash n } L_{m,n'}(v_{n'}),  \\
     s.t.  \sum_{n' \in \mathcal{N}(m)}h_{m,n'}^G v_{n'} = 0; \nonumber  \\
     L_{m,n}(v_n)\Longleftarrow L_{m,n}(v_n)- L_{m,n}(0);
\end{eqnarray*}

\item[S5:] A posteriori information computation: \\
\begin{equation*}
  \tilde{L}_n(v_n)=L_n(v_n) + \sum_{m \in \mathcal{M}(n)}
     L_{m,n}(v_n);
\end{equation*}
\item[S6:] Variable node processing : \\
\begin{eqnarray*}
  L_{n,m}(v_n) &=& \tilde{L}_n(v_n) - L_{m,n}(v_n),
 \end{eqnarray*}
and extrinsic information extraction : \\
\begin{eqnarray*}
  L_i(v_n) &=& \tilde{L}_n(v_n) - L_e(v_n);
\end{eqnarray*}
Go to step S2.
\end{enumerate}

The check node processing function can be computed recursively as
\begin{eqnarray}
\label{eqn:2}
     \bigoplus_{v_n\in V_m} L(v_n) = L(v_1) \oplus \left(\bigoplus_{v_n\in V_m\backslash v_1} L(v_n)\right),
\end{eqnarray}
where
\begin{equation}
   L(v_1)\oplus L(v_2): = L(v_1+v_2) =\left[L(v_1+v_2)=\alpha_i\right],
\end{equation}
and
\begin{eqnarray}
\label{eqn:2}
     L(v_1 + v_2 = \alpha_i)= \ln\left(\sum_{x\in GF(4)}e^{L(x)+L(\alpha_i-x)}\right) \nonumber \\
     -  \ln\left(\sum_{x\in GF(4)}e^{L(x)+L(-x)}\right).
\end{eqnarray}

\subsection{Comments}
The presented log-form version is, in essence, tutorial. In \cite{WubbenBPSK}, the G-SPA is proposed explicitly for decoding of LDPC coded modulation over synchronous MACs. Then, the authors in \cite{LuWeb} developed the G-SPA decoding of RA coded modulation over the symbol-asynchronous MAC.

Clearly, the presented Log-G-SPA for LDPC coded asynchronous relaying is the cascade of two sub-message-passing algorithms, which include the Log-BCJR algorithm ($\mathcal{B}_{MAC}$) for the asynchronous PNC and the Log-G-SPA for the two-user LDPC codes. In its current form, two sub-message-passing algorithms run one iteration in turn and then exchange information iteratively. In practice, the rates of convergence for these two sub-message-passing algorithms are different and  decoding of the LDPC code is often slower. Let us claim the outer iterations for $\mathcal{B}_{MAC}$, and the inner iterations for the Log-G-SPA decoding of the two-user LDPC codes. Hence, one can place more than one inner iterations for the Log-G-SPA decoding of the two-user LDPC codes for each outer iteration. This will be discussed in simulations.

\section{Frame-asynchronous Bi-directional Relaying}
\subsection{Channel Model}
\begin{figure}[htb]
   \centering
   \includegraphics[width=0.45\textwidth]{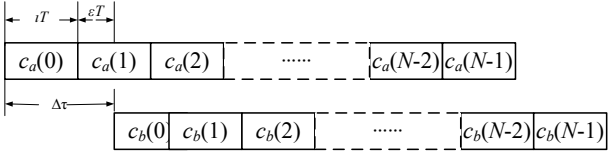}
   \caption{Frame asynchronism between sources A and B.}
   \label{fig:frAsyn}
\end{figure}

In section-II, the symbol-asynchronous multiple-access channel model is considered, where the relative relay $\delta=\frac{\tau_b-\tau_a}{T}$
is restricted to $0\leq \delta < 1$. In practice, the asynchronism between sources A and B cannot be controlled elegantly in this manner. The frame asynchronism should also be seriously considered. Hence, one can consider that $\delta = \iota +\epsilon$ as shown in Fig. \ref{fig:frAsyn}, where $\iota$ is a nonnegative integer number with $0\leq \iota \ll N$ and $\epsilon$ is a fractional value with $0\leq \epsilon < 1$. For $\iota \ll N$, it means that some control mechanisms for synchro-transmission between sources A and B are still required in the multiple-access phase, which, however, can be reasonably relaxed.

Let $\mathbf{c}_{ab}(k,\iota)=\left[\begin{array}{c}
c_a(k) \\
c_b(k-\iota)
\end{array}\right]$, and $\mathbf{r}(k)=\left[\begin{array}{c}
r_a(k) \\
r_b(k)
\end{array}\right]$. Then, the formulation (\ref{eq:W}) still holds with the form of
\begin{eqnarray}
 \label{eq:d4}
    \mathbf{r}(k) &=& \Psi \left(\{\mathbf{c}_{ab}(k-l,\iota)\}_{l=0}^L\right) + \mathbf{n}(k).
\end{eqnarray}

\subsection{Cyclic Codes for Combating the Frame-asynchronism}
For the frame-asynchronous bi-directional relaying, it is natural to ask the LDPC coding to have the following property
\begin{equation}
 \label{eq:cyc1}
    H (\mathbf{c}_a \oplus \mathbf{c}_b^{(\iota)}) = \mathbf{0},
\end{equation}
where $\mathbf{c}_b^{(\iota)}=(c_b(N-\iota+2),c_b(N-\iota+3),\cdots,c_b(N-1), c_b(0),c_b(1),\cdots,c_b(N-\iota+1))$ is the $\iota$-cyclic shift of $\mathbf{c}_b$.
It means that
\begin{equation}
 \label{eq:cyc2}
    H \mathbf{c}_b^{(\iota)} = \mathbf{0}.
\end{equation}
Hence, the LDPC code should be the cyclic code.

In this manner, one can still construct a virtual combined encoder which maps the messages generated by both sources A and B into the virtual codeword $\mathbf{c}_{ab}^\iota(D)=\left[c^{ab}_{0,\iota}(D),\cdots, c^{ab}_{N-1,\iota}(D)\right]$, where $c^{ab}_{n,\iota}(D)=c_a(n) + c_b(n-\iota) D$ with $n-\iota \triangleq  n-\iota \mod N$. Then, a virtual 4-ary LDPC coding scheme is again obtained with the codeword $\mathbf{c}_{ab}^\iota(D)$ satisfying
\begin{equation}
 \label{eq:cyc3}
   \sum_n h_{m,n}^G(D) c^{ab}_{n,\iota}(D) = 0  \mod (1+D+D^2)
\end{equation}
with a special constraint of $H^G=H$.

Let $\iota_{\max}$ be the integer part of the potential maximum delay, i.e., $-\iota_{\max} \le \iota \le \iota_{\max}$. For the received signal vector $\mathbf{r}_0^{N-1}$, the $\iota$ elements at both the head and the tail of  $\mathbf{r}_0^{N-1}$ can be thought as the interference part.
Let $\hat{\mathbf{r}}_0^{N-1} \triangleq [\mathbf{0}_{\iota_{\max}},\mathbf{r}_{\iota_{\max}}^{N-1-{\iota_{\max}}}, \mathbf{0}_{\iota_{\max}}]$, which can be seen as the worst case for extracting the $a$ $posterior$ information for $\mathbf{c}_{ab}(k,\iota)$.   With the log-BCJR algorithm for the asynchronous MAC, i.e., $\mathcal{B}_{MAC}\left(\hat{\mathbf{r}}_0^{N-1}, L_i(\mathbf{c}_{ab}(k,\iota)), L_e(\mathbf{c}_{ab}(k,\iota))\right)$, one can obtain the estimate of $\ln(\Pr(\mathbf{c}_{ab}(k,\iota)|\mathbf{r}_0^{N-1}))$. If we assume that the transmission of LDPC coded packets is continuous for both sources A and B, it is clear that both the head and the tail of each superimposed packet may be corrupted by the past and future packets. Indeed, the number of symbols corrupted is $2\iota$ for each LDPC frame, which results into the possible SNR loss of  $10\log10(\frac{N-2\iota}{N})$ in dB. If $\iota\ll N$, the SNR loss is minor.

With the cyclic LDPC codes applied at both sources A and B, the Log-G-SPA algorithm presented in Section-III can be again employed for getting the estimate of $\mathbf{c}_a \oplus \mathbf{c}_b^{(\iota)}$. However, the relative delay $\iota$ should be resolved.

\subsection{Resolving the Relative Delay $\iota$}
For the sources to correctly decode the messages, it is essential to resolve the unknown delay $\iota$. Here, we propose two potential mechanisms to solve this problem.
Let $\Upsilon^{(i)}(\mathbf{r}_0^{N-1})$ denote the output LLR vector after $i$-th message-passing decoding with the channel input vector $\mathbf{r}_0^{N-1}$.  Consider the case of continuous transmission from both sources A and B to the relay R. For the received signal vector $\mathbf{r}_0^{N-1}$, the $\iota$ elements at both the head and the tail of  $\mathbf{r}_0^{N-1}$ can be thought as the interference part.  In general, the interference part is detrimental for successful decoding of XOR codeword $(\mathbf{c}_a \oplus \mathbf{c}_b^{(\iota)})$. Hence, it is natural to replace the interference part with the all-zero vector. Based on this observation, we propose the following estimator
\begin{eqnarray}
  \label{eq:fr}
     \hat{\iota} = \max_{\iota \in U} \left|\Upsilon^{(i)}\left([\mathbf{0}_\iota,\mathbf{r}_{\iota}^{N-1-\iota},\mathbf{0}_\iota] \right)\right|.
\end{eqnarray}
In simulations, we found that this method does not work well as the contribution of the interference part is minor when the value of $\iota$ is small.

To correctly resolve the delay $\iota$ with high probability, we propose to employ cyclic redundancy check (CRC) codes to identify the message. CRCs are specifically designed to protect against common types of errors on communication channels, where they can provide quick and reasonable assurance of the integrity of messages delivered. In the consider scenarios, CRCs  are  appended to a message packet, which is further LDPC encoded for possible transmission at each source node.

With the Log-G-SPA decoding developed in Section-III, the relay decodes the received signal and can output two codewords $\hat{\mathbf{c}}_a$ and $\hat{\mathbf{c}}_b^{(\iota)}$. Then the codeword $\hat{\mathbf{c}}_b^{(\iota)}$ is cyclically-shifted to $l=-\iota_{\max},\cdots,\iota_{\max}$ positions and CRC-checking is performed to resolve the true value of $\iota$. Simulations show that this mechanism can work well.

If the JCNC scheme is employed, the thing is different. However, it is still possible to resolve the delay but this task has to be completed in the broadcasting phase. In this case, the relay decodes the received signal and output the codeword $\mathbf{c}_r=(\mathbf{c}_a \oplus \mathbf{c}_b^{(\iota)})$ with unknown $\iota$ at the MAC phase. Then the codeword $\mathbf{c}_r=(\mathbf{c}_a \oplus \mathbf{c}_b^{(\iota)})$ has been broadcasted to both sources. For the source A, the message-passing decoder is implemented to get an estimate of $\mathbf{c}_r$, which is further processed with its own codeword $\mathbf{c}_a$ to get the estimate of $\mathbf{c}_b^{(\iota)}$. Then, $\mathbf{c}_b^{(\iota)}$ is cyclically-shifted and CRC-checking is performed to identify the proper value of the shift $\iota$.  For the source B, the message-passing decoder is also implemented to get an estimate of $\mathbf{c}_r$. Then, $\mathbf{c}_b$ is cyclically-shifted to get the estimate of $\mathbf{c}_b^{(\iota)}$. Each copy of the cyclic-shift of $\mathbf{c}_b$ is employed to XOR the decoded $\mathbf{c}_r$ for obtaining the estimate of $\mathbf{c}_a$. Finally, CRC-checking  is again performed to identify the correctly-extracted codeword $\mathbf{c}_a$.

\section{Simulation Results}

In simulations, a square-root-raised-cosine filter is employed for pulse-shaping at the transmitter. The roll-off factor $\beta$ is chosen to be  $\beta=1$ for both sources A and B. As the bottleneck of the bi-directional relaying system lies in the processing capability of the relay node and its performance.
For joint LDPC and PNC scheme, it is the duty of the relay to reproduce the XOR of both source codewords. Hence, we mainly focus on the performance of the XOR codeword $\mathbf{c}_r$. The performance is closely related to the energy per bit and the received noise variance. The BPSK modulation is adopted.

Two LDPC codes are considered. The first one is a (3,6)-regular Mackay-Neal LDPC code with codewords of length $N=1008$\cite{MacKay}, and the second one is a cyclic LDPC code derived from finite-geometry codes with codeword length of $N=1365$ and information length $K=765$\cite{QHuang}. Let us denote the maximum number of outer iterations ($\mathcal{B}_{MAC}$) by $n_o=4$, and the maximum number of inner (LDPC) decoding iterations per outer iteration by $n_i$.

In simulations, we consider the normalized equal-power complex MAC channel, namely, $|h_a|=|h_b|=1$. This complex MAC channel is often characterized by the carrier-offset $\Delta \theta = \theta_b-\theta_a$, where $h_a=|h_a|e^{j\theta_a}$ and $h_b=|h_b|e^{j\theta_b}$ are complex variables. Throughout the simulations, the symbol misalignment with $\epsilon=0.5$ is considered.

\begin{figure}[htb]
   \centering
   \includegraphics[width=0.45\textwidth]{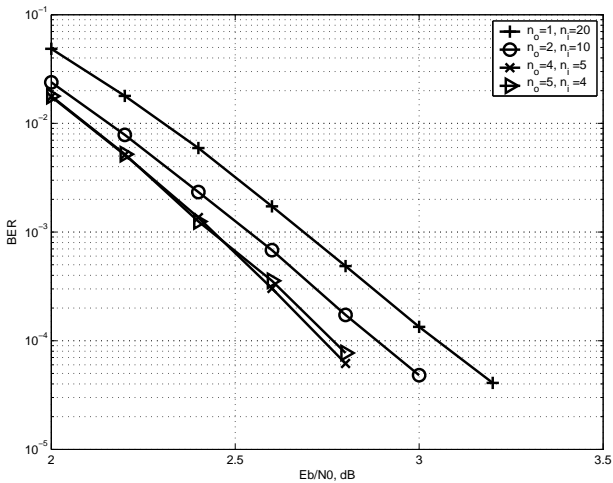}
   \caption{The effect of the number of the outer iterations for the Log-G-SPA decoding}.
   \label{fig:1}
\end{figure}

Firstly, the effect of $n_o$, the number of $\mathcal{B}_{MAC}$ iterations, on the system performance is investigated. The number of total inner LDPC decoding iterations is fixed to $n_i\cdot n_o=20$. For the (3,6)-regular Mackay-Neal LDPC code, the BER performance is shown in Fig. \ref{fig:1} for the normalized equal-power complex MAC channel with $\Delta \theta= \pi/4$. As shown, the exchange of information between $\mathcal{B}_{MAC}$ and the LDPC decoder can be beneficial for performance enhancement, and $n_o=4$ is enough for the considered case.

\begin{figure}[htb]
   \centering
   \includegraphics[width=0.45\textwidth]{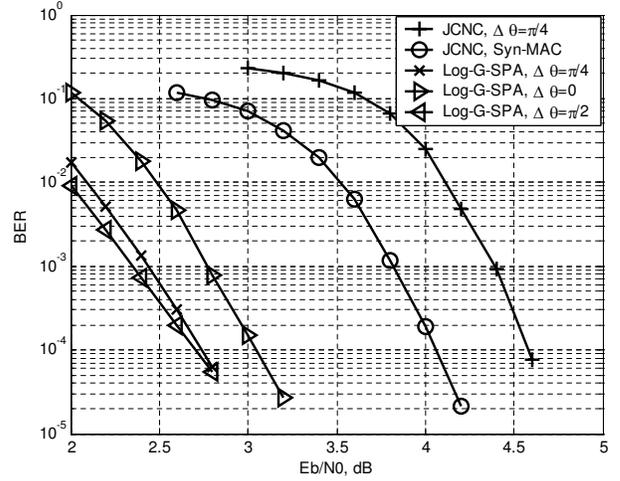}
   \caption{The bit-error-rate performance of the (1008,504) Mackey-Neal LDPC over Symbol-Asynchronous and Synchronous MACs with different $\Delta \theta$s.}
   \label{fig:2}
\end{figure}

Secondly, the BER performance of Log-G-SPA decoding with symbol-asynchronism is presented with various values of $\Delta \theta$, which is further compared to that of the JCNC decoding. It is shown in Fig. \ref{fig:2} that the Log-G-SPA performs significantly better than the JCNC, and its robustness to $\Delta \theta$ is also observed with symbol-asynchronism compared to the JCNC \cite{LuWeb}. For the JCNC decoding, the performance under the symbol-asynchronism is deteriorated compared to the case of the synchronous MAC (Syn-MAC).

\begin{figure}[htb]
   \centering
   \includegraphics[width=0.45\textwidth]{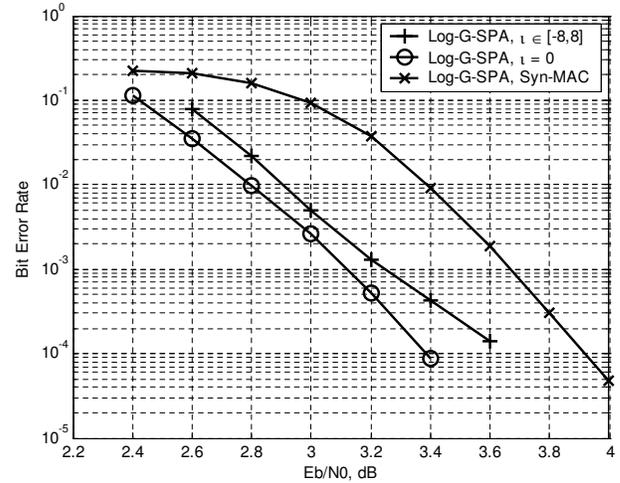}
   \caption{The bit-error-rate performance of the (1365,765) cyclic LDPC over Frame-Asynchronous and Synchronous MACs with $\Delta \theta = \frac{\pi}{4}$}.
   \label{fig:3}
\end{figure}

Finally, we investigate the effect of both frame and symbol misalignments on the system performance. Hence, the cyclic LDPC code is employed and the maximum number of iterations are set to $n_o=4, n_i=10$. For frame misalignment, $\iota$ is assumed to be randomly picked from $[-8,8]$. The BER performance is shown in Fig. \ref{fig:3} for the normalized equal-power complex MAC channel with $\Delta \theta= \pi/4$. As shown, the performance degradation due to the frame misalignment is less than 0.2 dB in this case. If the mechanism of (\ref{eq:fr}) is incorporated, the performance gap can be further narrowed but the complexity must be increased to about $2 \iota$ times of the original algorithm.   To resolve the relative delay $\iota$, the CRC-16 is appended to the source message of length $K_m=K-16=749$ for both sources A and B. In simulations, the resolution process is only initiated when the parity-checks of both LDPC codes are satisfied, i.e.,  the error-free case. We found that the resolution process always works and the correct value of $\iota$ can be found once the parity-checks of LDPC codes are satified.

\section{Conclusion and Future Work}
We have presented a general channel model for the asynchronous multiple-access channel with arbitrary pulse-shaping, typically encountered in bi-directional relaying. By evoking the WMF technique,  one can arrive at an equivalent vector ISI channel, which can be employed to develop the well-known BCJR algorithm for getting the optimal APPs.

We also present a formal Log-G-SPA decoding for the LDPC coded BPSK signalling over asynchronous MACs. To further combat the frame asynchronism, the cyclic LDPC codes, along with the CRC techniques, are proposed. Various proposed techniques can be well employed to overcome the asynchronism of the bi-directional relaying.

There are several issues to be explored in future. For either continuous or burst transmission of LDPC frames, the proposed methods can only work with limited misalignment in frames, i.e., a small value of $\iota$ relative to the codeword length. This method generally requires to some control mechanism for cooperative transmission between sources A and B. If no control mechanism is required, one must seriously consider the case of large values of $\iota$. For this case, one have to base several LDPC frames for decoding and the interference cancelation technique can be helpful for successful decoding and correct resolution of unknown delay.

The Log-G-SPA decoding has shown its performance advantage if both sources A and B employ the same LDPC code. However, it remains unknown how to design LDPC codes for maximizing the system performance with either synchronous or asynchronous multiple-access channels. Furthermore, compared to the Log-SPA decoding of non-binary LDPC codes,   the complexity of Log-G-SPA decoding for LDPC coded transmission over MACs is somewhat lower since the permutation step is not required. However, there are more work to do for its practical implementation.

\section*{Acknowledgment}
The authors wish to thank Dr. Qin Huang for providing the parity-check matrix of the $(1365,765)$ cyclic LDPC code. We also thank Dr. Ming Jiang for helpful discussions.


\end{document}